\documentclass[openacc]{rstransa}%%%%where rstrans is the template name

\begin{document}

%%%% Article title to be placed here
\title{Counterdiabatic driving for periodically driven open quantum systems}

\author{%%%% Author details
Kazutaka Takahashi}

%%%%%%%%% Insert author address here
\address{Institute of Innovative Research, Tokyo Institute of Technology,
Kanagawa 226--8503, Japan \\
Department of Physics Engineering, Faculty of Engineering,
Mie University, Mie 514--8507, Japan}

%%%% Subject entries to be placed here %%%%
\subject{quantum physics, quantum thermodynamics, quantum control}

%%%% Keyword entries to be placed here %%%%

%%%% Insert corresponding author and its email address}
%\corres{Kazutaka Takahashi\\
%\email{ktaka@phen.mie-u.ac.jp}}

%%%% Abstract text to be placed here %%%%%%%%%%%%
\begin{abstract}
We discuss dynamics of periodically-driven open quantum systems.
The time evolution of the quantum state is described by 
the quantum master equation and the form 
of the dissipator is chosen so that the instantaneous stationary state 
is given by the Gibbs distribution.
We find that the correlation between
the population part and the coherence part of the density operator 
is induced by an adiabatic gauge potential.
Although the introduction of the counterdiabatic term 
eliminates the correlation,
additional correlations prevent a convergence to the Gibbs distribution.
We study the performance of the control by the counterdiabatic term.
The system has three different scales and 
the performance strongly depends on the relations among their magnitudes.
\end{abstract}
%%%%%%%%%%%%%%%%%%%%%%%%%%%

%%%%%%%%%% Insert the texts which can accomdate on firstpage in the tag "fmtext" %%%%%

\begin{fmtext}
\end{fmtext}

%%%%%%%%%%%%%%% End of first page %%%%%%%%%%%%%%%%%%%%%

\maketitle

%%%%%%%%%%%%%%%%%%%%%%%%%%%%%%%%%%%%%%%%%%%%%%%%%%%%%%%%%%%%%%%%%%%%%%%%%%%
\section{Introduction}

When we drive the system dynamically, 
we observe a nontrivial change of the quantum state.
The generator of the time evolution is dependent on time and
the eigenstate basis of the generator is not useful to represent the state.
To describe such systems, the method of shortcuts to adiabaticity 
is shown to be useful and we can find many possible applications in 
literatures~\cite{DR03, DR05, Berry09, CRSCGM, STA13, STA19, Takahashi19}.

The application of the method to open systems with dissipation effects 
is an important question to be asked and, in fact, various studies have 
been discussed since the development of the method.
In open quantum systems, the notion of adiabaticity is not 
obvious~\cite{Sarandy05}.
For example, 
we can discuss a shortcut in decoherence-free subspaces~\cite{Wu17}.
We can also control the system by engineering 
the environments~\cite{Alipour20, Dupays20}.
Nonadiabatic effects in the quantum master equation
were also discussed~\cite{Dann18, Dann19}.
Since the dissipation effect is represented by a nonhermitian operator, 
the problem can be discussed in the context of 
nonhermitian systems~\cite{Ibanez11}.

There are many possible realizations of the coupling to the environment
and the application of the method depends on the situation that 
we want to describe.
To treat open quantum dynamics, 
we pay attention to the consistency with thermodynamical properties.
When the system is coupled to a thermal bath, 
the system relaxes to the Gibbs state.

In the present work, we operate the system periodically in time.
This operation is relevant 
when we study the performance of the heat engine for example.
Theoretically, the periodic operation of the system means 
that we treat time dependent Hamiltonian.
Then, when the system is coupled to the thermal bath,
thermalization behavior is not obvious.

When we slowly operate the system,
the system basically follows the instantaneous Gibbs state.
In this situation, we examine correlations between 
the population part and the coherence part of the density operator.
To treat the open quantum dynamics, we treat 
the Gorini--Kossakowski--Lindblad--Sudarshan (GKLS) 
equation~\cite{Gorini76, Lindblad76, Breuer02, Lidar19}.
The form of the dissipator is chosen so that 
the Gibbs state represents the instantaneous stationary solution.

We note that the adiabatic pucture is drastically changed 
when we consider the fast driving.
Then, the thermalization behavior is 
not obvious~\cite{DR14, LDM14, MIKU18}.
Although the method of the counterdiabatic driving in closed systems 
works irrespective of the speed of the driving, 
we show in the following that 
the counterdiabatic driving strongly depends on the speed.

The organization of the paper is as follows.
In section \ref{gkls}, we describe the GKLS equation in
a thermodynamically consistent form.
Then, in section \ref{popcoh}, 
we define the instantaneous eigenstate basis of the Hamiltonian 
to represent the density operator.
Based on a vector representation of the equation, 
we discuss correlations between 
the population part and the coherence part.
In section \ref{cd}, 
we introduce the counterdiabatic term 
and examine the property by using a simple two-state system.
The last section \ref{conc} is devoted to conclusions.

%%%%%%%%%%%%%%%%%%%%%%%%%%%%%%%%%%%%%%%%%%%%%%%%%%%%%%%%%%%%%%%%%%%%%%%%%%%
\section{Quantum master equation in a thermodynamically consistent form}
\label{gkls}

We consider a finite-dimensional quantum state represented 
by the density operator $\hat{\rho}(t)$.
The dimension of the Hilbert space is denoted by $N$.
The time evolution of the density operator is described by 
the GKLS equation 
\begin{align}
 \frac{\partial}{\partial t}\hat{\rho}(t)
 =-i[\hat{H}(\lambda(t)),\hat{\rho}(t)]+\hat{D}_\beta[\hat{\rho}(t)].
\end{align}
The dynamical property of the system is characterized by 
the Hamiltonian $\hat{H}(\lambda(t))$ 
and the dissipator $\hat{D}_\beta[\hat{\rho}(t)]$.

The system Hamiltonian $\hat{H}(\lambda)$ is parametrized by 
a set of time-dependent parameters 
$\lambda(t)=(\lambda_1(t),\lambda_2(t),\dots)$.
We consider the case where the system is driven periodically.
Then, the period is given by $T_0=2\pi/\omega$ 
and the parameters satisfy $\lambda(t+T_0)=\lambda(t)$. 

The system is coupled to a single external reservoir and the coupling 
is characterized by the dissipator $\hat{D}_\beta[\hat{\rho}(t)]$ 
where $\beta$ denotes the inverse temperature of the reservoir.
We use a thermodynamically consistent form of the dissipator 
so that the stationary state is given by 
the instantaneous Gibbs distribution
$\hat{\rho}_{\rm s}(t)\propto e^{-\beta\hat{H}(\lambda(t))}$.
In the long-time limit $t\gg T_0$ where 
the state is independent of the initial condition,  
we examine how the density operator deviates from the Gibbs distribution.

To obtain a thermodynamically consistent form of the dissipator, 
we write the system Hamiltonian $\hat{H}(\lambda(t))$ 
by using the spectral representation as
\begin{align}
 \hat{H}(\lambda)=\sum_{n=1}^N\epsilon_n(\lambda)|\epsilon_n(\lambda)\rangle
 \langle\epsilon_n(\lambda)|.
\end{align}
Here, $\{\epsilon_n(\lambda)\}_{n=1,2,\dots,N}$ and 
$\{|\epsilon_n(\lambda)\rangle\}_{n=1,2,\dots,N}$ 
represent a set of eigenvalues and 
the corresponding set of eigenstates respectively.
We assume for simplicity $\epsilon_m(\lambda)\ne\epsilon_n(\lambda)$ 
for $m\ne n$.
The eigenstates satisfy the standard orthonormal relation
$\langle\epsilon_m(\lambda)|\epsilon_n(\lambda)\rangle=\delta_{m,n}$
and the resolution of unity 
$\sum_{n=1}^N|\epsilon_n(\lambda)\rangle\langle\epsilon_n(\lambda)|=1$.

The explicit form of the dissipator is written by using 
the spectral representation as 
\begin{align}
 \hat{D}_\beta[\hat{\rho}] = \sum_{\alpha,\epsilon}\gamma_{\alpha}(\epsilon)
 \left[
 \hat{L}_{\alpha}^{\epsilon}\hat{\rho}(\hat{L}_{\alpha}^{\epsilon})^\dag 
 -\frac{1}{2}\left( 
 (\hat{L}_{\alpha}^{\epsilon})^\dag \hat{L}_{\alpha}^{\epsilon}
 \hat{\rho}+\hat{\rho}
 (\hat{L}_{\alpha}^{\epsilon})^\dag\hat{L}_{\alpha}^{\epsilon}
 \right)
 \right]. \label{dissipator}
\end{align}
$\hat{L}_{\alpha}^{\epsilon}$ represents a jump operator with 
energy eigenstate projections.
A Hermitian jump operator $\hat{L}_\alpha$ is microscopically 
introduced from the system-reservoir coupling Hamiltonian and 
$\hat{L}_{\alpha}^{\epsilon}$ is obtained as 
\begin{align}
 \hat{L}_{\alpha}^{\epsilon}=\sum_{m,n}\delta_{\epsilon,\epsilon_m-\epsilon_n}
 |\epsilon_n\rangle\langle \epsilon_n|
 \hat{L}_{\alpha}|\epsilon_m\rangle\langle \epsilon_m|.
\end{align}
This operator satisfies the relation
\begin{align}
 \hat{L}_{\alpha}^{-\epsilon} = (\hat{L}_{\alpha}^{\epsilon})^\dag.
\end{align}
We note that $\hat{L}_{\alpha}^{\epsilon}(\lambda)$ is 
interpreted as a lowering operator since it satisfies the commutation relation
\begin{align}
 [\hat{H},\hat{L}_{\alpha}^{\epsilon}] = -\epsilon\hat{L}_{\alpha}^{\epsilon}
\end{align}
at each time.
Correspondingly, $(\hat{L}_{\alpha}^{\epsilon})^\dag$ is interpreted 
as a raising operator.
The dissipator coupling $\gamma_\alpha(\epsilon)$ is microscopically 
introduced from a correlation function of an operator for the reservoir.
It is a nonnegative quantity and satisfies the 
Kubo--Martin--Schwinger condition~\cite{Kubo57, Martin59}
\begin{align}
 \gamma_{\alpha}(-\epsilon)=  e^{-\beta\epsilon}\gamma_{\alpha}(\epsilon). 
 \label{dbc}
\end{align}
In this setting, 
we obtain the relation $\hat{D}_\beta[\hat{\rho}_{\rm s}]=0$ where
$\hat{\rho}_{\rm s}$ represents the Gibbs state
\begin{align}
 \hat{\rho}_{\rm s}(t)=\frac{1}{Z(\lambda(t))}e^{-\beta\hat{H}(\lambda(t))}
 = \frac{1}{Z(\lambda(t))}\sum_{n=1}^N e^{-\beta\epsilon_n(\lambda(t))}
 |\epsilon_n(\lambda(t))\rangle\langle\epsilon_n(\lambda(t))|. \label{gibbs}
\end{align} 
The normalization $Z(\lambda)$ represents the partition function 
of the instantaneous Hamiltonian $\hat{H}(\lambda)$.

The point discussed in this section is that the dissipator is dependent 
not only on the state $\hat{\rho}(t)$ 
but also on the Hamiltonian $\hat{H}(t)$.
The use of the thermodynamically consistent form of the dissipator 
is crucial to find specific properties 
on the counterdiabatic driving in open quantum systems.

%%%%%%%%%%%%%%%%%%%%%%%%%%%%%%%%%%%%%%%%%%%%%%%%%%%%%%%%%%%%%%%%%%%%%%%%%%%
\section{Correlation between population and coherence parts}
\label{popcoh}

In the present work, we basically focus on a regime 
where the frequency $\omega$ is small enough 
compared to the other scales in the GKLS equation.
Then, we can use the expansion with respect to $\omega$.
It corresponds to the adiabatic approximation, 
though the term ``adiabatic'' is misleading for open systems treated 
in the present study.  
We examine how the population part and the coherence part 
of the density operator correlate with each other~\cite{Uzdin15}.

Generally, by using the diagonal basis of the density operator,
we can find that the dissipator consists of two parts~\cite{Funo19}.
They describe the population dynamics and the coherence dynamics.
Although this picture is useful to understand the structure of dynamics,
the decomposition of the equation is not so obvious.
We must solve the nonlinear problem which is usually a difficult task.

When the system is driven by slow-changing parameters, 
we can use the instantaneous eigenstates of the Hamiltonian.
In fact, in the zeroth order of the expansion 
with resppect to the time derivative of the parameters 
the density operator is given by the Gibbs state in Eq.~(\ref{gibbs}).

We generally write the density operator as
\begin{align}
 \hat{\rho}(t)=\sum_{mn}|\epsilon_m(\lambda(t))\rangle\rho_{mn}(t)
 \langle\epsilon_n(\lambda(t))|.
\end{align}
We note that the choice of the phase 
in the eigenstate is arbitrary in principle.
However, we can learn from the general theory of the adiabatic approximation
that the phase of the instantaneous eigenstate should be chosen 
so that the following relation holds:
\begin{align}
 \langle\epsilon_n(\lambda(t))|\partial_t|\epsilon_n(\lambda(t))\rangle=0. 
 \label{enn}
\end{align}
This choice is always possible by changing the state as
\begin{align}
 |\epsilon_n(\lambda(t))\rangle\to 
 |\epsilon_n(\lambda(t))\rangle\exp\left(-\int_0^t dt'\,
 \langle\epsilon_n(\lambda(t'))|\partial_{t'}
 |\epsilon_n(\lambda(t'))\rangle\right).
\end{align}

We write the equation of motion by using 
$\rho_{mn}(t)=\langle\epsilon_m(\lambda(t))|
\hat{\rho}(t)|\epsilon_n(\lambda(t))\rangle$.
We decompose the time derivative of the density operator as 
\begin{align}
 \partial_t\hat{\rho}(t)=\hat{\rho}'_0(t)+\hat{\rho}'_1(t).
\end{align}
The first term is written as 
\begin{align}
 & \hat{\rho}'_0(t)=\sum_{mn}|\epsilon_m(\lambda(t))\rangle\dot{\rho}_{mn}(t)
 \langle\epsilon_n(\lambda(t))|, 
\end{align}
where the dot symbol denotes the time derivative.
The second term arises when the eigenstate is dependent on $t$.
It can be written in a commutation-relation form
\begin{align}
 & \hat{\rho}'_1(t)=-i[\dot{\lambda}(t)\cdot\hat{A}(\lambda(t)),
 \hat{\rho}(t)], \\
 & \hat{A}(\lambda)= i\sum_{m,n}
 |\epsilon_m(\lambda)\rangle
 \langle\epsilon_m(\lambda)|\partial_\lambda |\epsilon_n(\lambda)\rangle
 \langle\epsilon_n(\lambda)|.
\end{align}
As we discuss in the next section, $\hat{A}$ represents 
the adiabatic gauge potential used in the counterdiabatic driving.

Using $\rho_{mn}(t)$, we rewrite the GKLS equation 
in a vector space~\cite{Mukamel99}.
We define the vector  
\begin{align}
 |\rho(t)\rangle =\left(\begin{array}{c} 
 |\rho_{\rm pop}(t)\rangle \\  |\rho_{\rm coh}(t)\rangle 
 \end{array}\right), 
\end{align}
where the set of diagonal components 
$|\rho_{\rm pop}(t)\rangle =(\rho_{11}(t),\rho_{22}(t),\dots,\rho_{NN}(t))^{\rm T}$
describes the population dynamics 
and the rest 
$|\rho_{\rm coh}(t)\rangle =(\rho_{12}(t),\rho_{13}(t),\dots,\rho_{N-1N}(t))^{\rm T}$
describes the coherence dynamics. 
The GKLS equation is written in a master equation like form as
\begin{align}
 \frac{\partial}{\partial t}|\rho(t)\rangle = {\cal K}(t)|\rho(t)\rangle,
\end{align}
where the matrix ${\cal K}$ takes a form 
\begin{align}
 {\cal K}(t)=\left(\begin{array}{cc} K(\lambda(t)) & 
 i\dot{\lambda}(t)\cdot A^{(12)}(\lambda(t)) \\
 i\dot{\lambda}(t)\cdot A^{(21)}(\lambda(t)) & 
 K_2(\lambda(t))-i\Delta(\lambda(t))
 +i\dot{\lambda}(t)\cdot A^{(2)}(\lambda(t)) \end{array}\right).
 \label{calk}
\end{align}
Each part is described below.

The matrix $K$ in the upper left block of ${\cal K}$ 
is decomposed as 
$K(\lambda)=\sum_{\alpha}K_{\alpha,\epsilon}^\epsilon(\lambda)$.
The offdiagonal component is given by 
\begin{align}
 (K_\alpha^{\epsilon}(\lambda))_{mn}=\gamma_\alpha(\epsilon(\lambda))
 |\langle\epsilon_m(\lambda)|\hat{L}_\alpha^\epsilon(\lambda)
 |\epsilon_n(\lambda)\rangle|^2, \label{K1}
\end{align}
where $m\ne n$.
The diagonal component $(K_\alpha^\epsilon(\lambda))_{nn}$
is determined by the relation 
$\sum_m (K_\alpha^\epsilon(\lambda))_{mn}=0$.
These properties show that $K$ is interpreted 
as a transition-rate matrix used in the classical master equation.
This property is reasonable since we obtain the classical result 
when we neglect contributions from the coherence part.
We can introduce the Gibbs state as the zero-eigenvalue state of $K$: 
$K(\lambda(t))|\rho_{\rm s}(t)\rangle=0$.

The offdiagonal blocks of  ${\cal K}$ induce correlations between
the population and coherence parts.
Their components are written as 
\begin{align}
 & i(\dot{\lambda}(t)\cdot A^{(12)}(\lambda(t)))_{m,kl}
 = -\delta_{ml}\langle\epsilon_m|\dot{\epsilon}_k\rangle
 +\delta_{mk}\langle\epsilon_l|\dot{\epsilon}_m\rangle, \label{a12}\\
 & i(\dot{\lambda}(t)\cdot A^{(21)}(\lambda(t)))_{kl,m}
 = \delta_{km}\langle\epsilon_k|\dot{\epsilon}_l\rangle
 -\delta_{lm}\langle\epsilon_k|\dot{\epsilon}_l\rangle. \label{a21}
% & i(\dot{\lambda}(t)\cdot A^{(2)}(\lambda(t)))_{mn,kl}
% = -\delta_{ln}\langle\epsilon_m|\dot{\epsilon}_k\rangle
% +\delta_{km}\langle\epsilon_l|\dot{\epsilon}_n\rangle,  
\end{align}

The lower right block of ${\cal K}$ consists of three parts.
$\Delta$ comes from the Hamiltonian and takes a diagonal form as 
\begin{align}
 \Delta_{mn,mn}(\lambda)=\epsilon_m(\lambda)-\epsilon_n(\lambda).
\end{align}
$A^{(2)}$ represents the adiabatic gauge potential and the explicit form 
is written in a similar way as $A^{(12)}$ and $A^{(21)}$.
$K_2$ comes from the dissipator and 
each component takes generally a complex value.

The equation for the population part is treated as 
a classical master equation.
The coupling to the coherence part 
by $\dot{\lambda}\cdot A^{(12)}$ induces a quantum effect.
When the time dependence of the Hamiltonian is weak, 
we can use the expansion with respect to the frequency.
At the zeroth order of the expansion, each part satisfies 
\begin{align}
 & K(\lambda(t))|\rho_{\rm pop}(t)\rangle =0, \\
 & (K_2(\lambda(t))-i\Delta(\lambda(t)))|\rho_{\rm coh}(t)\rangle =0. 
\end{align}
As we mentioned in the previous section, the solution of these equations 
represents the stationary Gibbs state, which means 
$|\rho_{\rm pop}(t)\rangle=|\rho_{\rm s}(t)\rangle$ 
and  $|\rho_{\rm coh}(t)\rangle=0$.
We note that $K$ is a square matrix and 
has the zero eigenvalue due to the property $\sum_mK_{mn}=0$.
We also see that $K_2-i\Delta$ is a square matrix and is basically invertible.

Up to the first-order of the expansion, 
the coherence part is written as 
\begin{align}
 |\rho_{\rm coh}(t)\rangle\sim 
 -i(K_2-i\Delta)^{-1}\dot{\lambda}\cdot A^{(21)}|\rho_{\rm s}(t)\rangle,
\end{align}
Then, the equation for the population part in the second-order accuracy 
is obtained as 
\begin{align}
 |\dot{\rho}_{\rm pop}(t)\rangle\sim \left(
 K+\dot{\lambda}\cdot A^{(12)}(K_2-i\Delta)^{-1}\dot{\lambda}\cdot A^{(21)}
 \right)|\rho_{\rm pop}(t)\rangle.
\end{align}
The second term represents the quantum correction 
to the classical master equation.

To find the explicit form of the solution, we expand the vectors 
$|\rho_{\rm pop}(t)\rangle=\sum_{k=0}^\infty|\rho_{\rm pop}^{(k)}(t)\rangle$ and 
$|\rho_{\rm coh}(t)\rangle=\sum_{k=0}^\infty|\rho_{\rm coh}^{(k)}(t)\rangle$.
The superscript denotes the order of the expansion.
We assume that $K$ is represented by the spectral representation 
\begin{align}
 K(\lambda)=\sum_n\Lambda_n(\lambda)|R_n(\lambda)\rangle\langle L_n(\lambda)|.
\end{align}
$\langle L_n|$ and $|R_n\rangle$ are 
the left eigenstate and the right eigenstate respectively.
They satisfy the orthonormal relation 
$\langle L_m(\lambda)|R_n(\lambda)\rangle=\delta_{m,n}$ and 
the resolution of unity 
$\sum_n |R_n(\lambda)\rangle\langle L_n(\lambda)|=1$.
We assign $n=0$ for the zero eigenvalue $\Lambda_0=0$.
Then, the stationary solution of the population part 
at the zeroth order is given by 
$|\rho_{\rm s}^{(0)}(t)\rangle=|R_0(\lambda(t))\rangle$.
Using the zeroth-order solution, we obtain the first-order contributions 
\begin{align}
 & |\rho_{\rm pop}^{(1)}(t)\rangle=\bar{K}^{-1}(\lambda(t))\partial_t
 |R_0(\lambda(t))\rangle, \\
 & |\rho_{\rm coh}^{(1)}(t)\rangle
 =-(K_2(\lambda(t))-i\Delta(\lambda(t)))^{-1}
 \dot{\lambda}(t)\cdot A^{(21)}(\lambda(t))
 |R_0(\lambda(t))\rangle,
\end{align}
where
\begin{align}
 \bar{K}^{-1}(\lambda)
 =\sum_{n(\ne 0)}\frac{1}{\Lambda_n(\lambda)}
 |R_n(\lambda)\rangle\langle L_n(\lambda)|.
\end{align}
The first-order term contribution $|\rho_{\rm pop}^{(1)}(t)\rangle$
is independent of the eigenstates of the Hamiltonian and  
is nonzero when the corresponding eigenvalues are dependent on $t$.
This form leads to a geometrical interpretation.
When the system is operated periodically, 
a heat current contribution from $|\rho_{\rm pop}^{(1)}\rangle$
in the adiabatic treatment  
is represented by a flux penetrating a closed surface 
in parameter space~\cite{Thouless83, Sinitsyn07}.
A similar interpretation is possible even if we improve 
the approximation~\cite{Takahashi20}.

%%%%%%%%%%%%%%%%%%%%%%%%%%%%%%%%%%%%%%%%%%%%%%%%%%%%%%%%%%%%%%%%%%%%%%%%%%%
\section{Counterdiabatic driving}
\label{cd}

%%%%%%%%%%%%%%%%%%%%%%%%%%%%%%%%%%%%%%%%%%%%%%%%%%%%%%%%%%%%%%%%%%%%%%%%%%%
\subsection{Counterdiabatic term}

The formal result in the previous section shows that 
the offdiagonal blocks of ${\cal K}$ are independent of the dissipator and 
the correlation between the population part and the coherence part 
is induced by the time dependent eigenstates of the Hamiltonian. 
The offdiagonal forms are represented by Eqs.~(\ref{a12}) and (\ref{a21}) 
and can be eliminated by changing the Hamiltonian 
$\hat{H}(\lambda(t))\to\hat{H}(\lambda(t))+\hat{H}_{\rm cd}(t)$, where 
$\hat{H}_{\rm cd}(t)=\dot{\lambda}(t)\cdot\hat{A}(\lambda(t))$.
This additional term is called the counterdiabatic term.
The cancellation of the nondiagonal terms in the adiabatic basis means that 
nonadiabatic transitions are prevented even if we drive 
the system rapidly~\cite{Berry09}.
The present form of the quantum master equation shows that
the same mechanism works even for open systems.
We can confirm easily that $A^{(12)}$, $A^{(21)}$, and $A^{(2)}$ are 
canceled out by introducing $\hat{H}_{\rm cd}(t)$.

However, in contrast to the closed systems, 
the introduction of the counterdiabatic term in the open systems 
does not give the decoupling of the population part and the coherence part.
The modification of the Hamiltonian leads to a different form of the dissipator.
The projection basis in the dissipator becomes the eigenstates 
of $\hat{H}(\lambda(t))+\hat{H}_{\rm cd}(t)$, not of $\hat{H}(\lambda(t))$.
As a result, ${\cal K}$ still has offdiagonal contributions.

By introducing the counterdiabatic term, we can write the density operator
up to the first order as 
\begin{align}
 & |\rho_{\rm pop}(t)\rangle
 \sim \left(1+\bar{K}^{-1}\partial_t\right)|\rho_{\rm s}(t)\rangle, \label{pop}\\
 & |\rho_{\rm coh}(t)\rangle
 \sim 
 -i\left[(K_2-i\Delta)^{-1} -(-i\Delta)^{-1}\right] 
 \dot{\lambda}\cdot A^{(21)}|\rho_{\rm s}(t)\rangle. \label{coh}
\end{align}
The population part is unchanged and the coherence part is modified 
so that $|\rho_{\rm coh}(t)\rangle\to 0$ in the absence of the dissipation.
This expression implies that the 
coherence contributions are partially suppressed 
by introducing the counterdiabatic term.
At least, the method works at the regime $|K_2|\ll |\Delta|$ where 
the dissipation effect is small.
On the other hand, we basically assume that the system follows
the Gibbs distribution at the slow-driving regime.
To obtain the Gibbs state, the dissipation effect is necessarily required.
Thus, the introduction of the counterdiabatic driving causes nontrivial 
effects in open systems and is not so obvious as we found in closed systems.

%%%%%%%%%%%%%%%%%%%%%%%%%%%%%%%%%%%%%%%%%%%%%%%%%%%%%%%%%%%%%%%%%%%%%%%%%%%
\subsection{Two-level Hamiltonian}

As a demonstration, we treat a simple two-level system 
in a slow-driving regime analytically.
The general form of the two-state Hamiltonian is written as 
\begin{align}
 & \hat{H}(\lambda(t))= \frac{h(t)}{2} n(t)\cdot\hat{\sigma}, \label{h0}\\
 & n(t)=\left(\sin\theta(t)\cos\varphi(t), \sin\theta(t)\sin\varphi(t), 
 \cos\theta(t)\right), 
\end{align}
where $\hat{\sigma}=(\hat{\sigma}^x,\hat{\sigma}^y,\hat{\sigma}^z)$
represents the set of Pauli operators, and 
$\lambda(t)=(h(t),\theta(t),\varphi(t))$ represents a set of 
periodic functions with the period $T_0=2\pi/\omega$.

The diagonal blocks of ${\cal K}$ are written as 
\begin{align}
 & K =\Gamma\left(\begin{array}{cc} -1 & e^{-\beta h} \\
 1 & -e^{-\beta h}  \end{array}\right), \\
 & K_2-i\Delta =
 \left(\begin{array}{cc} -\Gamma_2-ih & 0 \\
 0 & \Gamma_2+ih \end{array}\right),
\end{align}
where $\Gamma$ and $\Gamma_2$ are contributions from the dissipator.
They are positive quantities and their explicit forms are shown below.
Using the formula derived in the previous section, 
we obtain the density operator up to the first order of the expansion as 
\begin{align}
 & |\rho_{\rm pop}(t)\rangle \sim \frac{1}{2}\left(\begin{array}{c}
 1-\tanh\frac{\beta h(t)}{2} \\ 1+\tanh\frac{\beta h(t)}{2} \end{array}\right)
 +\frac{\partial_te^{-\beta h(t)}}{\Gamma(t) (1+e^{-\beta h(t)})^3}
 \left(\begin{array}{c} -1 \\ 1 \end{array}\right), \label{rhopop1}\\
 & |\rho_{\rm coh}(t)\rangle \sim \tanh\frac{\beta h(t)}{2}
 \left(\begin{array}{c} \langle\epsilon_2(t)|\dot{\epsilon}_1(t)\rangle^*
 \frac{1}{\Gamma_2(t)+ih(t)} \\ 
 \langle\epsilon_2(t)|\dot{\epsilon}_1(t)\rangle
 \frac{1}{\Gamma_2(t)-ih(t)} \end{array}\right).
\end{align}
The first term of $|\rho_{\rm pop}(t)\rangle$ represents the Gibbs state.

The explicit forms of $\Gamma$ and $\Gamma_2$ are basically 
dependent on the choice of the jump operator.
When we choose the jump operator $\hat{L}$ as the phase-damping form 
\begin{align}
 \hat{Z} = \left(\begin{array}{cc} 1 & 0 \\ 0 & -1 \end{array}\right),
 \label{lz}
\end{align}
$\Gamma$ and $\Gamma_2$ are given by 
\begin{align}
 & \Gamma(t) = \gamma_z(h(t))\sin^2\theta(t), \\
 & \Gamma_2(t) = \frac{1+e^{-\beta h(t)}}{2}
 \gamma_z(h(t))\sin^2\theta(t)
 +2\gamma_z(0)\cos^2\theta(t).
\end{align}
We can also choose the bit-flip form 
\begin{align}
 \hat{X} = \left(\begin{array}{cc} 0 & 1 \\ 1 & 0 \end{array}\right),
 \label{lx}
\end{align}
to obtain 
\begin{align}
 & \Gamma(t) = \gamma_x(h(t))(1-\sin^2\theta(t)\cos^2\varphi(t)), \\
 & \Gamma_2(t) = \frac{1+e^{-\beta h(t)}}{2}
 \gamma_x(h(t))(1-\sin^2\theta(t)\cos^2\varphi(t)) 
 +2\gamma_x(0)\sin^2\theta(t)\cos^2\varphi(t).
\end{align}
Since we consider the projections onto the energy eigenstates, 
the result is not sensitive to the choice of the jump operator, 
as we confirm below.

Now we consider the introduction of the counterdiabatic term.
For the general form of the two-level Hamiltonian in Eq.~(\ref{h0}),
the counterdiabatic term is calculated to give 
\begin{align}
 \hat{H}_{\rm cd}(t)=\frac{1}{2}n(t)\times\dot{n}(t)\cdot\hat{\sigma}.
\end{align}
This term is added to the Hamiltonian.
Then, the total Hamiltonian is written as  
\begin{align}
 \hat{H}(\lambda(t))+\hat{H}_{\rm cd}(t)
 =\sum_{n=1}^N E_n(t)|E_n(t)\rangle\langle E_n(t)|,
\end{align}
and the set of eigenstates $\{|E_n(t)\rangle\}$ is used to give 
the Liouville form.
To compare the result with that for the original system,
we need to use the common eigenstates.
We use the transformation 
\begin{align}
 \hat{\rho}(t)=\sum_{mnkl}|\epsilon_m\rangle \langle\epsilon_m|E_k\rangle
 \rho^{(E)}_{kl}\langle E_l|\epsilon_n\rangle\langle\epsilon_n|.
\end{align}
$\rho^{(E)}_{kl}$ is calculated 
from the method in the previous section 
in the basis of $\hat{H}+\hat{H}_{\rm cd}$.

$|E_n(t)\rangle$ is expressed as a function of 
$\lambda(t)=(h(t),\theta(t),\varphi(t))$ 
and their derivatives.
Up to the first order of the expansion, we find 
that the population part in Eq.~(\ref{rhopop1}) is unchanged and 
the coherence part is changed as 
\begin{align}
 |\rho_{\rm coh}(t)\rangle \sim \tanh\frac{\beta h(t)}{2}
 \left(\begin{array}{c} 
 \langle\epsilon_2(t)|\dot{\epsilon}_1(t)\rangle^*\left(
 \frac{1}{\Gamma_2(t)+ih(t)} -\frac{1}{ih(t)}\right) \\ 
 \langle\epsilon_2(t)|\dot{\epsilon}_1(t)\rangle
 \left(\frac{1}{\Gamma_2(t)-ih(t)}-\frac{1}{-ih(t)}\right)
 \end{array}\right).
\end{align}
This form is consistent with the general result in Eq.~(\ref{coh}).
The last term in each component represents 
the effect of the counterdiabatic driving.
This result shows that the coherence part is negligible when $\Gamma_2\ll h$.
As we know from the general theory of shortcuts to adiabaticity, 
the contribution from the counterdiabatic term eliminates 
coherence contribution of the unitary dynamics.
In the present case, the coherence part is dependent on the dissipator and 
the counterdiabatic term eliminates a part of the coherence contributions.
We also see from this expression that the scale of the original Hamiltonian
$\hat{H}(\lambda(t))$ must be large compared to that of the dissipator.
In the counterdiabatic driving for closed systems,
the scale of $\hat{H}$ does not affect the control accuracy.
The present analysis shows that the scale of $\hat{H}$ is important 
to obtain the ideal control of open systems. 

%%%%%%%%%%%%%%%%%%%%%%%%%%%%%%%%%%%%%%%%%%%%%%%%%%%%%%%%%%%%%%%%%%%%%%%%%%%
\subsection{Numerical analysis}

Since the analytical study is only restricted to slowly-driving regime,
we numerically solve the GKLS equation in a wider range of parameters.
As a measure of control, 
we introduce the trace distance between the time-evolved state 
and the instantaneous Gibbs state as 
\begin{align}
 d(t)=\frac{1}{2}{\rm Tr}\,|\hat{\rho}(t)-\hat{\rho}_{\rm s}(t)|.
\end{align}
This quantity goes to a small value 
when the counterdiabatic driving works well.

\begin{figure}[!ht]
%\centering\includegraphics[width=2.5in]{fig1.eps}
\centering\includegraphics[width=0.9\columnwidth]{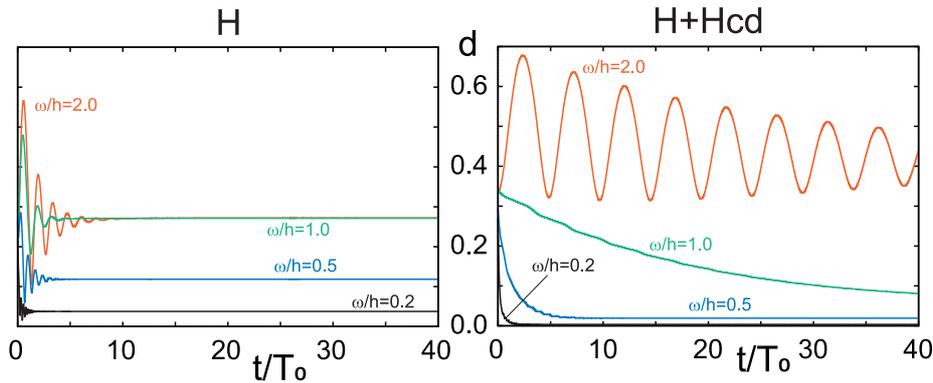}
\caption{
Time dependence of the trace distance.
We set $\beta h=1.0$ and the protocol is chosen as in Eq.~(\ref{p1}).
We use $\hat{L}=\hat{Z}$ and the coupling constant is set as 
$\gamma(h)/h=0.5$ and $\gamma(0)/h=0$.
The initial condition of the time evolution is given by 
$|\rho_{\rm pop}(0)\rangle=(0,1)^{\rm T}$ and $|\rho_{\rm coh}(0)\rangle=0$. 
The left figure denotes drivings without the counterdiabatic term 
and the right figure with the counterdiabatic term.
We plot results with several values of $\omega/h$.
}
\label{fig1}
\end{figure}

Equation (\ref{pop}) implies that 
the Gibbs distribution can be a good approximation when the eigenvalues of 
the Hamiltonian are independent of time.
Taking into account this property, we set $h$ as constant.
In Fig.~\ref{fig1}, we plot the result for the protocol 
\begin{align}
 h={\rm const.}, \qquad \theta=\frac{\pi}{4}, \qquad 
 \varphi(t)=\omega t. \label{p1}
\end{align}
We choose the jump operator as $\hat{L}=\hat{Z}$ in Eq.~(\ref{lz}).
The result shows that, after several periods of time, 
the density operator is close to the Gibbs state
when the system is driven slowly.
The behavior at the slow driving regime is further improved by 
introducing the counterdiabatic term.
On the other hand, 
the introduction of the counterdiabatic term can give 
instable behavior at the fast-driving regime.

\begin{figure}[!ht]
\centering\includegraphics[width=0.9\columnwidth]{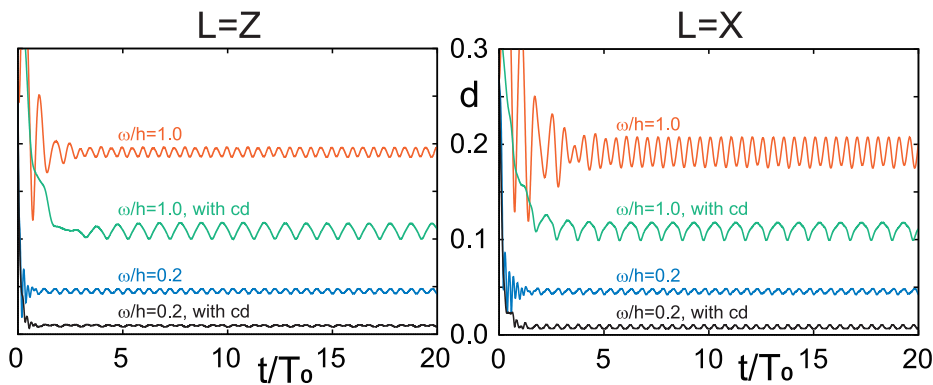}
\caption{
We set $\beta h=1.0$ and the protocol is chosen as in Eq.~(\ref{p2}).
We use the jump operator $\hat{Z}$ in the left figure and 
$\hat{X}$ in the right figure.
The coupling constant is set as 
$\gamma(h)/h=0.5$ and $\gamma(0)/h=0$.
The results for the original Hamiltonian $\hat{H}$ 
and the counterdiabatic term $\hat{H}+\hat{H}_{\rm cd}$ are plotted.
}
\label{fig2}
\end{figure}

We also plot the result for the protocol 
\begin{align}
 h={\rm const.}, \qquad \theta(t)=\frac{\pi}{2}\left(1-\frac{1}{5}\cos\omega t\right), 
 \qquad \varphi(t)=\omega t, \label{p2}
\end{align}
in Fig.~\ref{fig2}.
We choose the jump operator as 
$\hat{L}=\hat{Z}$ in the left panel of the figure 
and $\hat{L}=\hat{X}$ in the right panel.
We also find a similar behavior as that in Fig.~\ref{fig1}.
We see that the result is basically unchanged, 
at least in the slow-driving regime, 
even if we replace the jump operator to a different one.
This is due to the property that 
the projections to the energy eigenstates give a similar form 
of $\hat{L}^\epsilon$.

\begin{figure}[!ht]
\centering\includegraphics[width=0.9\columnwidth]{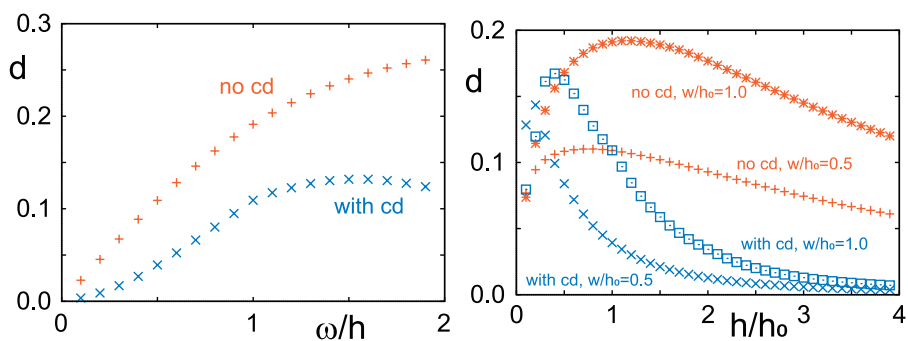}
\caption{
Trace distance as a function of $\omega$ (left) and of $h$ (right).
We put $\beta h_0=1.0$, $\hat{L}=\hat{Z}$, $\gamma(h)/h_0=0.5$, 
and $\gamma(0)/h_0=0$, 
where $h_0$ denotes the reference scale of $\hat{H}$.
}
\label{fig3}
\end{figure}

We also change other parameters.
In Fig.~\ref{fig3}, we plot the trace distance as a function of 
$\omega$ (left) and of $h$ (right).
We use the protocol in Eq.~(\ref{p2}) and 
the jump operator $\hat{L}=\hat{Z}$.
The results in Fig.~\ref{fig3} indicate that the control works well 
when the system satisfies the condition 
\begin{align}
 \omega \ll |\gamma| \ll |h|.
\end{align}
This behavior is consistent with the analytical result.

%%%%%%%%%%%%%%%%%%%%%%%%%%%%%%%%%%%%%%%%%%%%%%%%%%%%%%%%%%%%%%%%%%%%%%%%%%%
\section{Conclusion}
\label{conc}

We have discussed the dynamics of open quantum systems.
The dissipator in the quantum master equation 
takes a thermodynamically consistent form and
we introduce the eigenstate basis of the Hamiltonian 
to write the equation in a vector form.
This form is convenient to describe the slow-varying systems.
In this basis, the correlations between the population part 
and the coherence part are described by the adiabatic gauge potential.

The form of the quantum master equation implies that 
the correlation is reduced by introducing the counterdiabatic term
as we can demonstrate in closed systems.
However, we can still find a nontrivial correlation between 
the population and coherence parts since 
the dissipator is changed adaptively by the change of the Hamiltonian.
The addition of the counterdiabatic term works 
when the scale of the Hamiltonian is larger than that of the dissipator
and the frequency is smaller than other scales.
However, the former condition does not mean that 
the dissipator is unimportant.
To find the Gibbs state, the presence of the dissipator is important.
In the present setting, we operate the system periodically in time.
After operating the system for a long time, 
the system relaxes to a stationary distribution at each time.
The dissipator plays the crucial role to obtain such a behavior.

Our aim in this paper is to realize the instantaneous 
Gibbs distribution by operating the system periodically.
When we describe the heat engine 
by using the quantum master equation~\cite{Geva94, Bayona21},
the heat current is only dependent on the population part.
The friction effects are important to obtain the heat engine behavior 
and we must investigate the effect of the counterdiabatic term carefully.
This is one of interesting problems in the present formalism 
and will be discussed elsewhere.

\enlargethispage{20pt}

%\ethics{Insert ethics statement here if applicable.}

%\dataccess{Insert details of how to access any supporting data here.}

%\aucontribute{For manuscripts with two or more authors, insert details of the authors’ contributions here. This should take the form: 'AB caried out the experiments. CD performed the data analysis. EF conceived of and designed the study, and drafted the manuscript All auhtors read and approved the manuscript'.}

%\competing{Insert any competing interests here. If you have no competing interests please state 'The author(s) declare that they have no competing interests’.}

\funding{The author was supported by 
JSPS KAKENHI Grants No. JP20K03781 and No. JP20H01827. }

%\ack{Insert acknowledgment text here.}

%\disclaimer{Insert disclaimer text here if applicable.}

%%%%%%%%%% Insert bibliography here %%%%%%%%%%%%%%

\end{document}